\documentclass[10pt,aps,prl,showpacs,twocolumn,amsmath,amssymb,superscriptaddress]{revtex4-1}
\usepackage{amssymb,amsthm,amsfonts,amstext}
\usepackage{color}

\newcommand{\ket}[1]{\left| #1 \right\rangle}
\newcommand{\bra}[1]{\left\langle #1 \right|}

\providecommand{\ketbra}[2]{|#1\rangle\langle#2|}
\newcommand{\proj}[1]{\ket{#1}\!\bra{#1}}

\newcommand{\ev}[1]{\left\langle #1\right\rangle}
\DeclareMathOperator{\trace}{tr}
\DeclareMathOperator{\identity}{\mathbb{I}}

\definecolor{nred}{rgb}{0.7,0.2,0.2}
\definecolor{nblack}{rgb}{0,0,0}
\definecolor{nblue}{rgb}{0.2,0.2,0.7}
\definecolor{ngreen}{rgb}{0.2,0.6,0.2}

\begin{document}

\title{
Nonlocal multipartite correlations from local marginals
}

\author{Lars Erik W\"urflinger}
\email[Electronic address:]{lars.wurflinger@icfo.es}
\affiliation{ICFO-Institut de Ciencies Fotoniques, E-08860 Castelldefels, Barcelona, Spain}

\author{Jean-Daniel Bancal}
\affiliation{Group of Applied Physics, University of Geneva, CH-1211 Geneva 4, Switzerland}

\author{Antonio Ac\'in}
\affiliation{ICFO-Institut de Ciencies Fotoniques, E-08860 Castelldefels, Barcelona, Spain}
\affiliation{ICREA-Institucio Catalana de Recerca i Estudis Avan\c cats, Lluis Companys 23, 08010 Barcelona, Spain}

\author{Nicolas Gisin}
\affiliation{Group of Applied Physics, University of Geneva, CH-1211 Geneva 4, Switzerland}

\author{Tamas Vertesi}
\affiliation{Institute of Nuclear Research of the Hungarian Academy of Sciences, H-4001 Debrecen, P.O. Box 51, Hungary}

\date{\today}

\begin{abstract}
Understanding what can be inferred about a multi-particle quantum
system given only the knowledge of its subparts is a highly
non-trivial task. Clearly, if a global system does not contain information resource of some kind, 
nor do its subparts. For the case of entanglement as an information resource, it is known that the converse of this last statement is not true: some non-entangled reduced states are only compatible with global states which are entangled.
We extend this result to correlations and provide local marginal correlations that are only compatible with global genuinely tripartite nonlocal correlations.
Quantum nonlocality can thus be deduced
from the mere observation of local marginal correlations.
\end{abstract}

\maketitle
\section{\label{sec:intro}Introduction}

In contrast to classical systems, multipartite quantum systems can
be entangled and exhibit nonlocal correlations. Beyond their
fundamental interest, both properties are resources for quantum
information theory~\cite{Horodecki09,Barrett05}. It is thus a relevant question to understand
the types of quantum states and correlations that are possible in
composite quantum systems.

In a multipartite system, every subset of parties constitutes a
proper system in itself. The fact that these subsystems describe
parts of the same total system requires them to satisfy some
\emph{compatibility conditions}. For instance, a bipartite quantum
state $\varrho_{AB}$ is compatible with a tripartite state
$\varrho_{ABC}$ if and only if
$\varrho_{AB}=\trace_C(\varrho_{ABC})$. While it is
straightforward to check whether some reduced states are compatible
with a given global state, the question becomes much subtler when
the global state is unknown and one is interested in knowing
whether there exists a quantum state compatible with the given
marginals. Finding the conditions for compatibility among
reduced quantum states is known as the \emph{quantum
marginal problem} \cite{Linden2002,Higuchi,Hall,Klyachko2004}.
It is the quantum counterpart of the classical marginal problem, which is concerned with the compatibility of marginal probability
distributions.

The quantum marginal problem is trivial in the bipartite case: two
reduced states, $\varrho_A$ and $\varrho_B$, are always compatible
with the product bipartite quantum state
$\varrho_{AB}=\varrho_A\otimes\varrho_B$. However, the situation
becomes more interesting when more than two parties are
involved. For instance, it is well known that if two parties share
a maximally entangled state, then any tripartite quantum
state compatible with it must be such that the third party is
uncorrelated to the first two. This phenomenon is known as the monogamy of
entanglement~\cite{monogamyEnt, monogamy2} and implies that a
maximally entangled state $\ket{\phi^+}_{AB}$ is incompatible with
any correlated state $\rho_{AC}$ or $\rho_{BC}$. A similar
property, known as the monogamy of nonlocality, is displayed by nonlocal correlations~\cite{Barrett05}. Parts of a system can thus constrain the set of possible full systems in ways that show up in other parts of the same system.

In this work we are interested in the question of what can be
inferred about the correlations of a global state given only the
knowledge of some of its subparts. It is clear that if subparts of
a system display entanglement or nonlocality, so does the global
system. However, is the converse also true? For the case of
entanglement it is known that the answer to this question is
negative: there are separable states of two qubits that are only
compatible with entangled multipartite states~\cite{Toth1, Toth2}.
To show this, the authors of~\cite{Toth1, Toth2} used spin-squeezing-inequalities to detect
entanglement and found entangled multi-qubit states whose reduced
two-qubit states are separable. As the entanglement criteria they use only
rely on two-body correlations, this demonstrates the existence of
non-entangled reduced states that are only compatible with
entangled global states.

Here we pose a similar question in the context of no-signaling
correlations, where one deals with the raw correlations of
classical inputs and outputs described by a joint conditional
probability distribution. Therefore, one does not assume the whole
Hilbert space formalism of quantum mechanics but just the validity
of the no-signalling principle. Our goal, then, is to see whether
there are local marginal correlations that are only compatible
with multiparite nonlocal correlations. We show that this is
indeed the case and that, similarly to what happens with entanglement, nonlocality of multipartite correlations can be certified from marginal correlations that admit a local description.
We further provide a quantum state and
corresponding measurements that exhibit this type of correlations.
In this case we also demonstrate that the nonlocality present in
the full correlations can be genuinely
multipartite~\cite{pironio,gennonlocal}.
Concerning the question of certifying entanglement from separable marginals, we further provide new examples of separable reduced states that are only compatible with an entangled global state. Our findings show how the compatibility conditions lead to non-trivial
results even when acting on a priori useless marginals: it is
possible to witness the presence of useful correlations in the
global system from useless reduced states.

\section{\label{sec:correlations}Nonlocality from local marginals}
Quantum nonlocality represents a quantum property inequivalent to
entanglement. In the paradigm of device-independent quantum
information processing, nonlocality has been identified as an
alternative resource for quantum information protocols, necessary
for instance for secure key distribution~\cite{diqkd} or
randomness generation~\cite{dirng}. The corresponding scenario
consists of different distant observers that can input a classical
setting $x_i$ into his part of the system and obtains an output
$a_i$. The correlations of the inputs and outputs are encapsulated
in the joint conditional probability distribution $P(a_1\ldots\
a_N|x_1\ldots x_N)$ that denotes the probability of obtaining the
outputs $a_1,\ldots,a_N$ when inputs $x_1,\ldots, x_N$ are used.

In what follows, we consider a tripartite scenario where each
party can choose from two different inputs, denoted by $0$ and
$1$, and obtain two different outputs, denoted by $-1$ and $+1$,
that is $x,y,z\in \lbrace 0,1\rbrace$ and $a,b,c \in \lbrace -1,1
\rbrace$. It is useful to consider the following parametrization
of the probabilities
\begin{equation}
 \begin{aligned}
    P(abc|xyz)=&\frac{1}{8}\left[ 1 + a \ev{A_x} + b\ev{B_y} + c\ev{C_z}\right.\\
          & + ab\ev{A_xB_y}+ac\ev{A_xC_z} +bc \ev{B_yC_z}\\
&+\left.abc\ev{A_xB_yC_z} \right],
 \end{aligned}
\end{equation}
where $\ev{A_x}=P(a=1|x)-P(a=-1|x)$ is the expectation value for
the outcome of the first party $A$ given input $x$,
$\ev{A_xB_y}=P(ab=1|xy)-P(ab=-1|xy)$ is the expectation value for
the product of the outcomes of $A$ and $B$ given the inputs $x$
and $y$, and so on.

Given the fact that entanglement can be deduced from the observation of separable reduced states only \cite{Toth1, Toth2}, it seems
natural to ask whether one can infer that some tripartite
correlations are nonlocal, only from observation of local
bipartite marginals. To answer this question in the affirmative
one needs to find three local bipartite non-signaling
distributions $P_{AB},P_{AC},P_{BC}$ such that any tripartite
non-signalling distribution $P_{ABC}$ compatible with them is
nonlocal. Being compatible in this context means that one must
have
\begin{align}
\label{eq:pab}
 \sum_c P_{ABC}(abc|xyz)&=P_{AB}(ab|xy)\\
\label{eq:pac}
 \sum_b P_{ABC}(abc|xyz)&=P_{AC}(ac|xz)\\
\label{eq:pbc}
\sum_a P_{ABC}(abc|xyz)&=P_{BC}(bc|yz),
\end{align}
where the left hand sides are defined independently of the third
input as $P_{ABC}$ is assumed to be non-signaling. In what follows
we provide several examples of distributions satisfying these
requirements.

In the first example, we fix the one-party expectation values as
\begin{equation}
\label{eq:oneparty}
 \ev{A_x}=\ev{B_y}=\ev{C_z}=\frac{1}{3}, \qquad x,y,z \in\lbrace0,1\rbrace
\end{equation}
and the two-party expectation values as
\begin{equation}
\label{eq:twoparty}
\ev{A_xB_y}=\ev{A_xC_y}=\ev{B_xC_y}=\left\lbrace\begin{array}{rl}
1 &\text{if }  x=y=0,\\
-\frac{1}{3} & \text{otherwise.}
                                    \end{array}\right.
\end{equation}
These values define the three bipartite marginals univocally. One can check
that these bipartite correlations are local, as they satisfy all
possible permutations of the Clauser-Horne-Shimony-Holt (CHSH)
inequality~\cite{chsh}, which is the only relevant Bell inequality
for two parties having binary inputs and outputs~\cite{fine}.

However, only one tripartite non-signaling distribution has
\eqref{eq:oneparty} and \eqref{eq:twoparty} as its marginals. To see this, consider any tripartite
non-signaling distribution $P_{ABC}$ that is compatible with the
given marginals. The positivity constraints $P_{ABC}(abc|xyz)\geq
0$ together with the fixed values for the one- and two-party
expectation values lead to lower bounds on $\ev{A_xB_yC_z}$ and
$-\ev{A_xB_yC_z}$ that ultimately only allow for the assignment
\begin{equation}
\label{eq:threeparty}
 \ev{A_xB_yC_z}=\left\lbrace\begin{array}{rl}
\frac{1}{3} &\text{if }  x+y+z\in \lbrace0,1\rbrace,\\
-1 & \text{otherwise.}
                                    \end{array}\right.
\end{equation}
Equations \eqref{eq:oneparty}-\eqref{eq:threeparty} define an
extremal point of the tripartite non-signaling polytope, the box
number 29 in the classification of~\cite{tripartiteBoxes}. This
point is genuinely nonlocal as it violates a Svetlichny-Bell
inequality~\cite{tripartiteBoxes,svetlichny}. Thus we found some bipartite
correlations that are local, but only compatible with (unique)
genuinely tripartite nonlocal correlations.

While this first example answers our original question, it is not
entirely satisfactory, as no measurements on a quantum system can
achieve all bipartite correlations
\eqref{eq:oneparty}-\eqref{eq:twoparty} at the same time. Indeed, the only possible extension of
these correlations, namely box 29 in~\cite{tripartiteBoxes},
violates the ``Guess-Your-Neighbor-Input" inequality~\cite{gyni},
which is satisfied by quantum correlations. Let us thus provide a
general characterization of marginals that are only compatible
with nonlocal probability distributions. To this end, consider the set $\Pi$ of bipartite marginals with binary inputs and outputs, which result from a tripartite local and non-signalling probability distribution,
\begin{equation}
\label{eq:polytope}
\begin{split}
\Pi=\lbrace &(P_{AB},P_{AC},P_{BC})| \exists P_{ABC} \\ &\text{local s.t. \eqref{eq:pab},\eqref{eq:pac},\eqref{eq:pbc} hold}\rbrace.
\end{split}
\end{equation}
Clearly, the set $\Pi$ is convex and has a finite number of
extreme points. It is then a polytope and can be described by a
finite number of inequalities that only involve the marginal
correlations $P_{AB},P_{AC},P_{BC}$. If the bipartite marginals of
some tripartite non-signaling correlations violate any of these
inequalities, then they cannot be compatible with a local
tripartite distribution. Thus, any extension of these marginals to
a tripartite non-signaling distribution must be nonlocal. On the other hand, if some bipartite correlations satisfy all the inequalities that define $\Pi$, then they are necessarily compatible with some tripartite local correlations.

Similarly, one can check whether some marginals are compatible
with genuinely tripartite nonlocal correlations by considering the
polytope
\begin{equation}
\label{eq:polytope2}
\begin{split}
\Pi'=\lbrace &(P_{AB},P_{AC},P_{BC})| \exists P_{ABC} \\
&\text{bi-local s.t. \eqref{eq:pab},\eqref{eq:pac},\eqref{eq:pbc}
hold}\rbrace.
\end{split}
\end{equation}
Here we consider the definition of bilocality given in
Refs.~\cite{gennonlocal,pironio}, which solves some
inconsistencies of the original definition of bilocality by
Svetlichny~\cite{svetlichny}. Since the constraints of the
polytope $\Pi'$ are strictly weaker than those of $\Pi$, one has
$\Pi\subset\Pi'$. Any inequality satisfied by $\Pi'$ is thus also
a valid inequality for $\Pi$.

An example of inequality satisfied by $\Pi'$ (and $\Pi$) is:
\begin{equation}
\begin{split}
  \label{eq:ineq2}
-&\ev{A_0(1+B_0 +B_1 +C_0)}\\
-&\ev{A_1(1+B_0 +C_0 +C_1)}\\
-&\ev{B_0 +C_0 +B_0 C_0 +B_1C_1} \leq 4.
\end{split}
\end{equation}
Violation of this inequality implies that the correlations
compatible with the given marginals must be genuinely tripartite
nonlocal. The inequality \eqref{eq:ineq2} can be violated by
measuring the noisy W state $\varrho_{\text{W}}\left(p\right)$ for
$p>0.9548$, where
\begin{equation}
\label{eq:3wstate}
 \varrho_{\text{W}}\left(p\right) = p \proj{W} + \frac{1-p}{8}\identity,
\end{equation}
with $\ket{W}= \frac{1}{\sqrt{3}}\left(\ket{001} +\ket{010}
+\ket{100}\right)$ and $0\leq p \leq 1$. The corresponding
measurement settings are
\begin{equation}
 \begin{array}{ll}
  A_0 = \cos \alpha \sigma_z + \sin \alpha \sigma_x , \qquad  & A_1 =   \cos \alpha \sigma_z - \sin \alpha \sigma_x \\
 B_0=-\sigma_z, &B_1= \cos \beta \sigma_z + \sin \beta \sigma_x\\
C_0 = -\sigma_z,&C_1= \cos \beta \sigma_z - \sin \beta \sigma_x
 \end{array}
\end{equation}
and $\alpha=3.6241$ and $\beta=2.0221$. The reduced states of two
parties of $\varrho_{\text{W}}\left(p\right)$ are all equal and
have the form
\begin{equation}
\label{eq:redstate3}
 \varrho_{\text{red}}\left(p\right)= \frac{2p}{3}\proj{\psi^+} + \frac{p}{3}\proj{00} + \frac{1-p}{4}\identity,
\end{equation}
where $\ket{\psi^+}= 1/ \sqrt{2}(\ket{01}+\ket{10})$. Since these
reduced states satisfy the Horodecki criterion for the violation
of the CHSH inequality~\cite{horodeckicrit} for every $0\leq p
\leq 1$, any pair of two-outcome measurements on $\varrho_W(p)$ is
necessarily local. Thus we have obtained an example of local
quantum marginal correlations which are only compatible with
genuine tripartite nonlocal correlations.

\section{\label{sec:entanglement}Entanglement from separable marginals}


Regarding the problem of entanglement detection from separable marginals, note that the global state of a system is known to be generally determinable from its marginals, if one has the promise that the global state is pure \cite{Linden2002}.
Indeed, consider the bipartite marginals $\varrho_{AB}=\varrho_{AC}=\varrho_{BC}=\varrho=1/2 (\proj{00} + \proj{11})$. If the global state of the systems is pure, it follows from its Schmidt-decomposition that it must be the Greenberger-Horne-Zeilinger (GHZ) state $\ket{GHZ}=1/\sqrt{2}(\ket{000} +e^{i\phi}\ket{111})$.  While these bipartite marginals are separable, the GHZ state is entangled and, thus, observation of separable marginals can only be compatible with an entangled pure state. 


Now, if the global state is not assumed to be pure, then the above analysis
immediately fails. For instance, the reduced states of the GHZ
state are also compatible with the three-party mixed state
$\varrho_{ABC} = 1/2(\proj{000}+\proj{111})$, which is separable.
Thus, observation of these marginals without further knowledge on
the full state does not guarantee entanglement in the whole
system. Actually, this result applies to every graph state: for
any such state there is always a separable state that has the
same two-body reductions~\cite{Gittsovich2010a}. So no criterion
relying on two-particle correlations can detect graph-state
entanglement.

However, as mentioned before, it was shown that there are
separable two-qubit states that are only compatible with an
entangled global state \cite{Toth1, Toth2}. Here, we present
further examples of this feature involving the reduced states of
three-qubit states. The starting point for our investigation is
again a noisy W state.
The reduced states \eqref{eq:redstate3} are separable for $0\leq p
\leq p_\text{sep} =3/ \left(1 + 2 \sqrt{5}\right)$. We are
interested to see if there exists a value of $p$ with $p \leq
p_\text{sep}$ such that every three-qubit state compatible with
these reductions must be entangled.

To do that, we need to look for the maximal value of $p$ such that
every three-qubit state having $\varrho_\text{red}\left(p\right)$
as its reductions is \emph{not} entangled. For simplicity, let us
relax this last constraint, allowing the three-qubit state to have
a positive partial transposition (PPT) instead of being
separable~\cite{ppt}. After this relaxation, the maximal value of
$p$ corresponds to the solution $p^\star$ of the following
instance of semi-definite program (SDP):
\begin{equation}
\label{eq:primal3}
\begin{aligned}
p^\star =  &\; {\text{maximize}}
& & p \\
&\ \ \ \ \varrho,p&&\\
& \text{subject to}
& & \varrho \succeq 0,\\
&
&& \trace_X \varrho = \varrho_{\text{red}}\left(p\right)\;\text{for}\; X = A,B,C\\
&
&& \varrho^{T_X}\succeq 0\; \text{for} \;X=A,B,C.
\end{aligned}
\end{equation}
Note that the normalization condition $\trace(\varrho)=1$ is
ensured by the constraints on the bipartite marginals $\trace_X
\varrho$.

By constructing the dual to the previous problem, it is possible
to prove that the solution of~\eqref{eq:primal3} is $p^\star=3/(2
+ \sqrt{17})\simeq 0.4899$ (see Appendix). Therefore, the reduced
states~\eqref{eq:redstate3} with $p^\star<p\leq p_\text{sep}$
certify the presence of entanglement in the global state despite
being separable.

 \begin{table}[t]
 \centering
 \begin{center}
  \begin{tabular}{l|ccccc}
  $n$         & $3$ & $4$ & $5$ & $6$ &$7$\\
\hline
  $p^\star$       & $0.4899$ & $0.6180$ & $0.7464$ & $0.8279$ &$0.8787$  \\
  $p_\mathrm{sep}$& $0.5482$ & $0.7071$ & $0.8050$ & $0.8640$ & $ 0.9009$
  \end{tabular}
 \end{center}

 \caption{Values for separability of the reduced two-party states of the noisy W state,
 $p_\mathrm{sep}$, and for the solution to the SDP problem\eqref{eq:primal3}, $p^\star$,
 for different number of parties.}
 \label{tab:npartiesresult}
\end{table}
The above considerations can be generalised to the case of more
than three parties. Starting from the noisy W state of $n$ qubits
we found a similar behaviour: one can choose separable two-party
states that are only compatible with an entangled global state of
$n$ qubits. The value of $p_\mathrm{sep}$ for which the twoparty
reduced states become separable reads $p_\mathrm{sep}= n/(4-n
+2\sqrt{n^2 -4 n +8})$, while solving the corresponding SDPs
yields a value for $p^\star$. Table \ref{tab:npartiesresult}
summarises our results for $n\leq7$.

\section{\label{sec:conclusions}Conclusions}
To conclude, we have demonstrated how the compatibility constraints
among marginal distributions allow one to certify the presence of nonlocal correlations in a global state from marginals that allow a local description. In particular, we have provided examples of
local bipartite marginals that are only compatible with nonlocal
probability distributions, and even with genuinely tripartite nonlocal distributions. This result reveals that local models reproducing some (local) bipartite marginal correlations can be fundamentally incompatible with each other, since the full correlations representing their joint behavior admit no such model.

Furthermore, for the case of entanglement we have presented a collection of three separable two-qubit-states that are only compatible with an entangled tripartite state. From a general viewpoint, our work proves how compatibility constraints lead to non-trivial results even when acting on separable or local states.

\section{Acknowledgements}
We thank Geza Toth, Otfried G\"uhne, Yeong-Cherng Liang and Marco
Piani for useful discussions. This work was supported by the ERC
starting grant PERCENT, the CHIST-ERA DIQIP project, the European
EU FP7 Q-Essence and QCS projects, the Spanish FIS2010-14830
project, the Swiss NCCRs QP and QSIT, and the European ERC-AG
QORE.

\section{Appendix}
This appendix provides details on the solution of the SDP from main text.  Defining $M = 2/3 \proj{\psi^+} +1/3 \proj{00} -1/4 \identity$, the corresponding dual problem of \eqref{eq:primal3} can be written as
\begin{equation}
\label{eq:dual3}
\begin{aligned}
d^\star =  &\; {\text{minimize}}
& & \frac{1}{4}\trace \left(N_A +N_B+N_C \right) \\
&\ \ N_X,Q_X&&\\
& \text{subject to}
& & Q_X \preceq 0 \;\text{for}\;X=A,B,C\\
&
&& \trace \left(M\left(N_A +N_B+N_C \right)\right) =-1\\
& && \sum_{X} \identity_X\otimes N_X + Q_X^{T_X}\succeq 0.
\end{aligned}
\end{equation}
where $N_X$ are $4\times 4$-matrices and $Q_X$ are $8\times
8$-matrices; the expression $\identity_X \otimes N_X$ denotes the
operator that acts as the identity on particle $X$ and as $N_X$ on
the rest.

From weak duality one always has $d^\star \geq p^\star$. Every feasible point for the primal problem gives a lower bound $p'\leq p^\star$ and every dual feasible point gives an upper bound $d'\geq d^\star$. The following choice of the variables $\varrho$, $N_X$, $Q_X$ satisfy all the constraints of \eqref{eq:primal3} and \eqref{eq:dual3}, while yielding the same bounds $d'=p' =3/(2 + \sqrt{17})\simeq 0.4899$. Thus, we have $p^\star = d^\star = 3/(2 + \sqrt{17})$.

\begin{equation}
\begin{split}
 \varrho =&  \frac{p^\star}{2} \left(\proj{W} +\proj{\overline{W}}\right)+ \frac{3(1-p^\star)}{4}\sigma\\
  & +\frac{p^\star}{6}\proj{000} +\frac{3-5p^\star}{12}\proj{111}
\end{split}
\end{equation}
with $\sigma = 1/3 \left(\proj{001}+\proj{010}+\proj{100}\right)$ and $\ket{\overline{W}} = 1/\sqrt{3}\left(\ket{011} +\ket{101}+\ket{110}\right)$,

\begin{equation}
\begin{split}
N_X =\ & (1+\frac{5}{3\sqrt{17}})\frac{p^\star}{2}\proj{00}\\
&+ (1-\sqrt{17})\frac{p^\star}{12}(\proj{01}+\proj{10}) \\
&- (1+\frac{11}{\sqrt{17}}) \frac{p^\star}{6}(\ketbra{01}{10}\\
&+ \ketbra{10}{01})  + 2(\frac{1}{3}+\frac{1}{\sqrt{17}})p^\star\proj{11}
\end{split}
\end{equation}
for $X=A,B,C$, and

\begin{equation}
\begin{aligned}
Q_A& =\\ &(1+\frac{5}{3\sqrt{17}})\frac{p^\star}{4}(-\proj{000}+\ketbra{000}{110}\\
& +\ketbra{000}{101}  + h.c.)\\
&-(\frac{1}{3}-\frac{1}{\sqrt{17}})p^\star(\ket{001}+\ket{010})(\bra{001}+\bra{010})\\
&+\frac{4}{3\sqrt{17}}p^\star(\ketbra{001}{111}+\ketbra{010}{111}+h.c.)\\
&-(\frac{3}{5}-\frac{1}{3\sqrt{17}})\frac{p^\star}{2}(\proj{101}+\proj{110})\\
&+(\frac{1}{5}-\frac{7}{3\sqrt{17}})\frac{p^\star}{4}(\ketbra{101}{110}+\ketbra{110}{101})\\
&-2(\frac{1}{3}+\frac{1}{\sqrt{17}})p^\star\proj{111}
\end{aligned}
\end{equation}
where $h.c.$ stands for hermitian conjugate. $Q_B$ and $Q_C$ are equal to $Q_A$ after permutating the parties so that $B$ or $C$ take the role of $A$.


\end{document}